\documentclass{article}

\usepackage{microtype}
\usepackage{graphicx}
\usepackage{subfigure}
\usepackage{booktabs} % for professional tables

\usepackage{hyperref}

\usepackage[accepted]{icml2023}

\usepackage{amsmath}
\usepackage{amssymb}
\usepackage{mathtools}
\usepackage{amsthm}

\usepackage[capitalize,noabbrev]{cleveref}

\usepackage{amsmath}
\usepackage{amssymb}
\usepackage{mathtools}
\usepackage{amsthm}
\usepackage[capitalize,noabbrev]{cleveref}
\usepackage{color,amssymb,amsmath,epsfig,enumerate,multicol,bm,multirow,subfigure}
\usepackage{xcolor}
\newcommand{\revision}[1]{{#1}}
\newcommand{\ve}[1]{\boldsymbol{\mathbf{#1}}}

\usepackage[textsize=tiny]{todonotes}

\begin{document}

\twocolumn[
\icmltitle{Self-supervised Neural Factor Analysis for Disentangling 
Utterance-level Speech Representations}



\icmlsetsymbol{equal}{*}

\begin{icmlauthorlist}
\icmlauthor{Weiwei Lin}{equal,eie}
\icmlauthor{Chenhang He}{equal,comp}
\icmlauthor{Man-Wai Mak}{eie}
\icmlauthor{Youzhi Tu}{eie}
\end{icmlauthorlist}

\icmlaffiliation{eie}{Dept. of Electronic and Information Engineering,
The Hong Kong Polytechnic University, Hong Kong SAR, China.}
\icmlaffiliation{comp}{Dept. of Computing,
The Hong Kong Polytechnic University, Hong Kong SAR, China.}

\icmlcorrespondingauthor{Weiwei Lin}{weiwei.lin@connect.polyu.hk}

\icmlkeywords{self-supervised learning, factor analysis, speech recognition, speaker recognition, emotion recognition, language identification}

\vskip 0.3in
]



\printAffiliationsAndNotice{\icmlEqualContribution} % otherwise use the standard text.

\begin{abstract}
    Self-supervised learning (SSL) speech models such as wav2vec and HuBERT have demonstrated state-of-the-art performance on automatic speech recognition (ASR) and proved to be extremely useful in low label-resource settings. However, the success of SSL models has yet to transfer to utterance-level tasks such as speaker, emotion, and language recognition, which still require supervised fine-tuning of the SSL models to obtain good performance. We argue that the problem is caused by the lack of disentangled representations and an utterance-level learning objective for these tasks.  Inspired by how HuBERT uses clustering to discover hidden acoustic units, we formulate a factor analysis (FA) model that uses the discovered hidden acoustic units to align the SSL features. The underlying utterance-level representations are disentangled from the content of speech using probabilistic inference on the aligned features. Furthermore, the variational lower bound derived from the FA model provides an utterance-level objective, allowing error gradients to be backpropagated to the Transformer layers to learn highly discriminative acoustic units. When used in conjunction with HuBERT's masked prediction training, our models outperform the current best model, WavLM, on all utterance-level non-semantic tasks on the SUPERB benchmark with only 20\% of labeled data.
    \end{abstract}
    
    \section{Introduction}
    Supervised learning has driven the development of speech technologies for two decades. However, annotating speech data is considerably more challenging than other modalities. For example, automatic speech recognition (ASR) and language identification require linguistic knowledge. For speaker and emotion recognition, label ambiguity and human error are hard to avoid. 
    \begin{figure*}[!htp]
        \begin{center}
            \includegraphics[width=\textwidth]{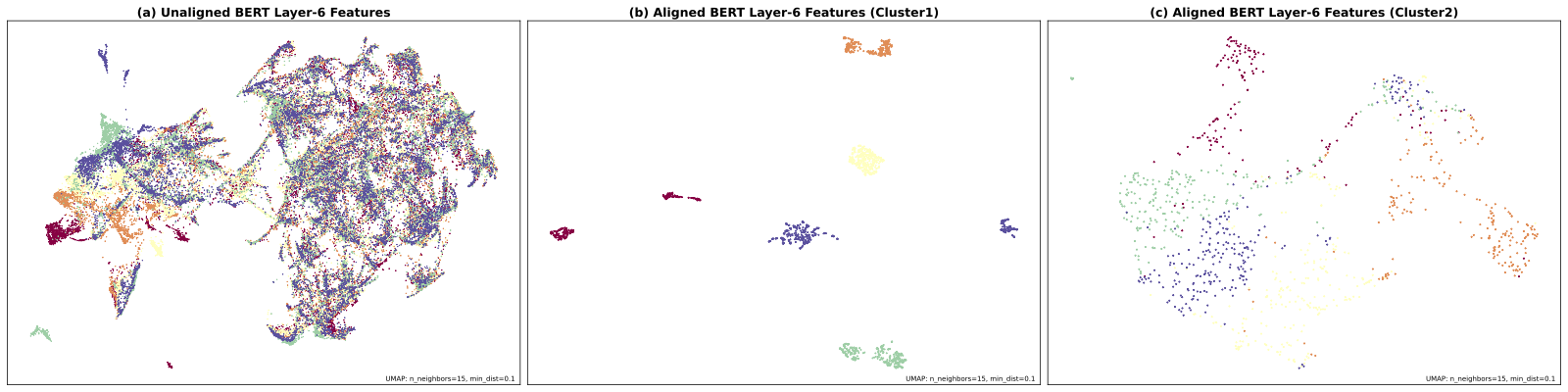}
                    \caption{Scatter plots of UMAP embeddings of Transformer features from HuBERT. Different colors represent different speakers. ``Aligned'' means that the frames were aligned using K-means.}
            \label{fig:umap}
        \end{center}
        \end{figure*}
    Self-supervised learning (SSL) promises a prospect of learning without labeled datasets. SSL speech models such as wav2vec \cite{wa2vec,wav2vec2} and HuBERT \cite{hubert} have profoundly changed the research landscape of ASR. By training on a large amount of unlabeled speech to learn a general representation and then fine-tuning with a small amount of labeled data, SSL models demonstrated state-of-the-art performance and proved to be very resource efficient in low label-resource settings \cite{hubert,wav2vec2}. 
    
    The success of wav2vec and HuBERT attracts researchers to apply SSL to other speech tasks \cite{wang2021fine}.
    For this purpose, Speech processing Universal PERformance Benchmark
     (SUPERB) for SSL models was proposed in \cite{superb}. 
    The tasks include content-based classifications, such as ASR, phoneme recognition, and intent classification, and utterance-level discriminative tasks, such as speaker recognition, diarization, and emotion recognition. SUPERB focuses on reusability of SSL features. Thus all tasks must share the same SSL model. Only the classification heads are learned using labeled data for a specific task. This encourages learning task-agnostic features for downstream tasks. \revision{Recently, a  NOn-Semantic Speech benchmark (NOSS) that specifically designed for utterance-level tasks was proposed in \cite{shor2020towards}. Using a triplet-loss unsupervised objective, they were able to exceeds the state-of-the-art performance on a number of transfer learning tasks.
    }
    
    Although it has been shown that SSL features can outperform hand-crafted features for almost all tasks \cite{superb} under the SUPERB protocols, the performance of supervised downstream models are still far behind the fully supervised or find-tuned models in utterance-level tasks, suggesting that directly using the SSL features to train the downstream models is not enough. 
    Besides, the labeled datasets in these tasks are considerably large. Using SSL models with little labeled data has yet to be explored for these tasks.
    % 2023 01/22
    This has led us to search for a more appropriate representation and an utterance-level self-supervised learning objective for these tasks.
    
    But, can an SSL model trained for frame-wise discrimination benefits utterance-level discrimination? We believe so. As shown in \cite{lei2014novel}, a DNN trained for phoneme classification can be used for training a powerful speaker verification system. 
    The key is in frame alignments. Averaging frame-level features cannot produce a good utterance representation because content variations within an utterance is too structural to be treated as Gaussian. 
    To demonstrate this, we randomly selected 200 recordings from 5 speakers in the LibriSpeech \cite{librispeech} test set and extracted speech features from the sixth Transformer layer of a HuBERT model. The UMAP \cite{umap} embeddings of the features are plotted in Figure~\ref{fig:umap}(a). Different colors in the figure represent different speakers. We cannot see any apparent speaker clusters in Figure~\ref{fig:umap}(a). If the content variations within an utterance are Gaussian, we should see blob-like speaker clusters. 
    One way to reduce content variations is to align frames according to phoneme-like units.
    However, the existing frame aligners either require supervised learning such as phoneme classification DNNs \cite{lei2014novel} or not amenable to stochastic gradient descent training such as Gaussian mixture models (GMM). 
    Inspired by HuBERT's use of K-means to discover hidden acoustic units, we propose aligning the frames using K-means. To this end, we trained a K-means model with 100 clusters on the LibriSpeech training set and used it to label the test set recordings. Then, we randomly selected two K-means clusters and only kept the frames assigned to these two clusters. The results are presented in Figures~\ref{fig:umap}(b) and (c). As we can see, the speaker clusters are clearly revealed with the help of K-means alignments.
    
    Specifically, we propose using the offline K-means model in HuBERT training to align the speech features. K-means is conceptually simple and amenable to the mini-batch training \cite{sculley2010web}. During HuBERT training, the K-means model is updated iteratively, which means the aligners can be gradually improved as well. With the K-means aligned features, we then decompose the utterance-level variations into a set of cluster-dependent loading matrices and a compact utterance-level vector. The utterance-level representation can be extracted using probabilistic inference on the aligned features.
    Finally, instead of using the EM algorithm to train the FA model as in many traditional FA approaches \cite{ivector}, we derived an utterance-level learning objective using the variational lower bound of the data likelihood. This allows gradients to be back-propagated to the Transformer layers to learn more discriminative acoustic features. Our experiments show that this objective can significantly improve the performance of SSL models on utterance-level tasks.
    %%%%%%%%%%%%%%%%%%%%%%%%%%%%%
    \begin{figure*}[!htp]
        \includegraphics[width=\textwidth]{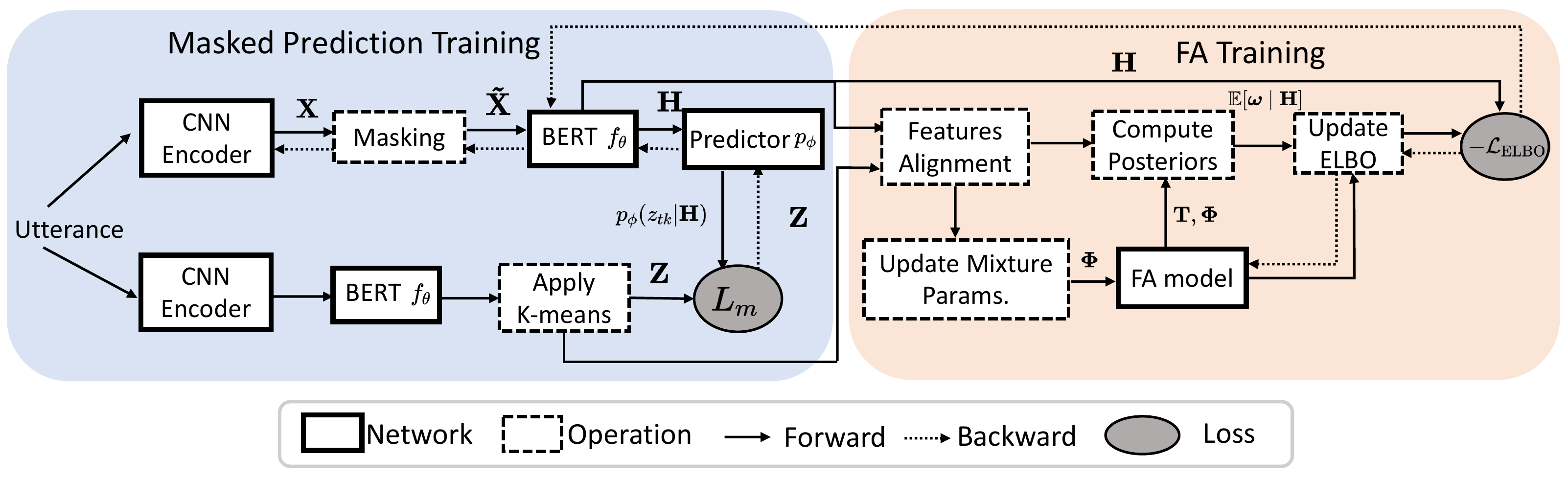}
        \caption{Training of the HuBERT variant of our neural factor analysis model. The dashed arrows represent gradient pathways. For the details of the learning algorithm, the reader may refer to Algorithm~\ref{alg:join}.}
    %	Note that there is no gradient (dashed arrow) flown back to the teacher model.}
        \label{fig:NFAM_training}
    \end{figure*}
    \section{Related Work}
    {\bf Self-supervised Learning for Speech}
    The majority of SSL approaches rely on pretext tasks, tasks that are not necessarily the direct objective but learning them can capture a high-level structure in the data \cite{bert,simclr,jagsaw}. In the speech community, some early attempts used multiple tasks as the learning pretexts \cite{paste,paste+}.
    An increasingly popular pretext is to use a context encoder to encode information about past frames to predict or reconstruct future frames, as pioneered by contrastive predictive coding (CPC) \cite{cpc}. This line of work includes wav2vec \cite{wa2vec}, which encodes raw waveform to perform frame differentiation, and autoregressive predictive coding \cite{apc} which uses an autoregressive model to predict future frames. Some researchers found that it is helpful to perform the frame discrimination on quantized representations \cite{vq-wav2vec,decoar}. Later, Transformers were used to encode both future and past contexts to perform frame discrimination, as in wav2vec 2.0 \cite{wav2vec2} and Mockingjay \cite{Mockingjay}.
     
    More recently, the Hidden-Unit BERT (HuBERT) was proposed for self-supervised speech representation learning \cite{hubert}. Different from explicit frame-wise discrimination in wav2vec and its variants, HuBERT is trained to perform masked prediction of pseudo labels given by an inferior HuBERT model from the previous optimization step. Later, multi-layer masked prediction losses were added to the intermediate layers of HuBERT to further strengthen the representation \cite{hubert2}. In ContentVec \cite{contectvec}, the authors improved HuBERT's performance for content-related tasks by disentangling speaker information from content information using voice conversion units. WavLM \cite{wavlm}, on the other hand, was proposed to improve both content-related tasks and utterance-level tasks by adding utterance mixing during training and gated relative position bias to the Transformer.
    %%%%%%%%%%%%%%%%%%%%%%%%%%%%
    
    {\bf Factor Analysis}
    Factor analysis (FA) and probabilistic models in general have wide applications in machine learning \cite{bishop2006pattern,murphy2012machine}. Before the advent of deep learning, there had been several successes of FA models in speaker verification, face recognition, and ECG signal classification, including joint-factor analysis \cite{JFA}, probabilistic linear discriminative analysis \cite{PLDA}, and most famously i-vector \cite{ivector}. 
    The FA models generally assume that there is a latent variable responsible for generating the observation vectors. Different relationships between the observation vectors and the latent variable result in different FA models, such as one-to-one mapping between the observation and the latent variable in probabilistic PCA and many observations to one latent variable in i-vector and JFA. Noticeably most of these FA models are applied to raw input or hand-craft features such as natural images or mel-frequency cepstral coefficients (MFCCs). 
    One exception is PLDA in speaker verification, which is applied to neural speaker embeddings or i-vectors. 
    
    {\bf Utterance-level Speech Tasks}
    Utterance-level speech tasks include speaker recognition \cite{svoverview}, emotion recognition \cite{emotion-overview}, and language identification \cite{langid-overview}. They are an important part of intelligent speech systems. Besides their respective applications, they are essential for semantic and generative tasks like ASR and text-to-speech (TTS) synthesis. For example, multilingual ASR and speech translation often require language identification as the first step \cite{whisper}. Multi-speaker TTS and voice conversion systems rely on speaker recognition models to extract speaker information \cite{multispk-tts,autovc}. 
    Solving these utterance-level tasks often involves different model architectures and domain knowledge.
    \section{Methodology}
    In this section, we will introduce our neural factor analysis (NFA) in the context of HuBERT. \revision{NFA aims to disentangle utterance-level information such as speaker identity, emotional state, and language from frame-wise content information such as phonemes.} Figure \ref{fig:NFAM_training} shows the training procedure of the HuBERT variant of our NFA model. 
    The learning objective we are about to derive can be used in any SSL model, such as wav2vec and its variants, as long as frame assignments are provided.
    NFA can learn various utterance-level representations, such as speaker identities, emotion states, and language categories. We will refer to them as utterance-level identities in the remaining paper.
    \subsection{HuBERT}
    Consider an acoustic sequence ${\bf X}$ of $T$ frames. We denote ${\cal M}\subset \{1,\ldots,T\}$ as the index set indicating the frames in ${\bf X}$ to be masked. 
    %Suppose ${\cal M} \subset[1,\hdots, T]$ is the indices to be masked for a $T$ length acoustic sequence $\ve{O}$. 
    Define $\tilde{\ve{X}}=\text{mask}(\ve{X}, {\cal M})$ as the masked version of $\ve{X}$, where the masked $\ve{x}_t$ $(t \in {\cal M})$ is replaced by a mask embedding. 
    The BERT encoder $f_{\ve{\theta}}(.)$ takes as input the masked sequence $\tilde{\ve{X}}$ and outputs a feature sequence $\ve{H} =[\ve{h}_1, \hdots, \ve{h}_T]$. 
    % 2023/01/23 2:30 pm notation is not correct
    \revision{Let us introduce a $K$-dimensional binary random variable $\mathbf{y}_t$ for frame $t$ having a 1-of- $K$ representation, where $y_{t k} \in 0,1$ and $\sum_k y_{t k}=1$. Denote the output of the predictor as $q_\phi\left(y_{t k} \mid \mathbf{H}\right)$ . Given the target distribution for the masked frames $p\left(y_{t k}\right)$, the cross-entropy can be computed as:
    \begin{equation}
        L_m(\mathbf{H}, \mathcal{M})=-\sum_{t \in \mathcal{M}} \sum_k p\left(y_{t k}\right) \log q_\phi\left(y_{t k} \mid \mathbf{H}\right)
    \end{equation}
    However, we do not have access to the target distribution $p\left(y_{t k}\right)$. HuBERT solves this problem by iterative clustering to obtain the frame label $z_{t k}$ as a surrogate for $p\left(y_{t k}\right)$, where $z_{t k} \in 0,1$ and $\sum_k z_{t k}=1$. With the frame label $z_{t k}$, the cross-entropy loss can be re-written as:
    \begin{equation}
    L_m(\mathbf{H}, \mathbf{Z}, \mathcal{M})=-\sum_{t \in \mathcal{M}} \sum_k z_{t k} \log q_\phi\left(y_{t k} \mid \mathbf{H}\right)
    \end{equation}}
    At first, the cluster assignments are obtained by running {\it K}-means clustering on MFCCs. Then the model is updated by minimizing the masked prediction loss.
    New cluster assignments are obtained by running {\it K}-means on the updated features at the Transformer layer. The learning process then proceeds with new cluster assignments $\{\ve{z}_t\}$. The masked prediction and cluster refinement are performed iteratively. 
    The blue area in Figure~\ref{fig:NFAM_training} illustrates HuBERT's masked prediction training.
    \begin{algorithm*}[tb]
       \caption{Training procedure of the proposed NFA model}
       \label{alg:join}
    \begin{algorithmic}
       \STATE \textbf{Initialize}:
             BERT parameters $\ve{\theta}$, 
             predictor  parameters $\ve{\phi}$,
             Loading matrix $\ve{T}$, 
             Initial cluster labels $\{\ve{Z}^{i}\}_{i=1}^{I}$.
      \FOR{$n \gets 0$ to $N$ iterations}
       \STATE \textbf{Input}: CNN encoder output $\{\ve{X}^{i}\}_{i=1}^{I}$, masking index set ${\cal M}$. 
      \IF{{\it n} $>$ 0}
       \STATE Run {\it K}-means on the BERT features to obtain frame labels $\{\ve{Z}^{i}\}_{i=1}^{I}$
       \ENDIF
       \STATE Use the alignments $\{\ve{Z}^{i}\}_{i=1}^{I}$ and Transformer features $\{\ve{H}^{i}\}_{i=1}^{I}$ to compute cluster parameters $\ve{\Phi}$.
        %%%%%%%%%%%%%%%%%%%%
        \FOR{$i \gets 1$ to $I$}
    %  	\STATE{S}
      \STATE  {\it \# Forward Pass}
        \STATE Mask the encoder output $\tilde{\ve{X}}^{i} = \text{mask}(\ve{X}^{i}, {\cal M})$.
        \STATE Calculate BERT output ${\mathbf{H}^{i}} = f_{\ve{\theta}}(\tilde{\ve{X}}^{i})$
        \STATE Calculate the posteriors of the latent factor (Eq.~\ref{eq:e-step}) and use them to update the ELBO $\mathcal{L}_{\text{ELBO}} \left(\ve{H}^{i};\ve{T}\right)$ (Eq.~\ref{eq:elbo_loss}).
    \STATE {\it \#  Backward Pass}
        \STATE Calculate the gradients on cross entropy loss $L_{m}(\ve{H}^{i}, \ve{Z}^{i}, {\cal M})$.
        \STATE Calculate the ELBO gradients with respect to $\ve{T}$ (Eq.~\ref{eq:grad-to-T}).
        \STATE Calculate the ELBO gradients with respect to the Transformer parameters $\ve{\theta}$ (Eq.~\ref{eq:grad-to-theta})).
        \STATE Update $\ve{\theta}$, $\ve{\phi}$, and $\ve{T}$ using gradient descent.    
        \ENDFOR
     
      \ENDFOR
     \STATE \textbf{Return} $\ve{\theta}$, $\ve{T}$
    \end{algorithmic}
    \end{algorithm*}
    %%%%%%%%%%%%%%%%%%%%%%%%%
    %%%%%%%%%%%%%%%%%%%%%%%%%
    % 2023 01/23 611pm
    \subsection{Utterance-level Representation Learning via Neural Factor Analysis}
    Figure~\ref{fig:umap} shows that the K-means alignments can reveal meaningful speaker information. One simple way to obtain the utterance-level representation is to average the aligned frames in each cluster and concatenate the results. The probabilistic model for such approach can be written as follows:
    \begin{equation}
    \mathbf{h}_{t}^{i}  \sim \sum_{k=1}^{K} z_{tk}^{i} \mathcal{N}\left(
    \boldsymbol{\mu}_{k} + \ve{w}_{k}^{i}, \ve{\Sigma}_k
    \right),
    \label{eq:ivector_fa}
    \end{equation}
    where $\mathbf{h}_{t}^{i}$ is the Transformer layer features from the utterance $i$, $z_{tk}^{i}\in\{0,1\}$ is the frame label assigned by K-means, $\boldsymbol{\mu}_{k}$ is the $k$-th cluster center, $\ve{\Sigma}_k$ is the covariance matrix of the $k$-th cluster, and $\ve{w}_{k}^{i}$ is the utterance identity in the $k$-th cluster. The concatenation of $\ve{w}_k^{i}$, i.e. $[\ve{w}_1^{i},\hdots \ve{w}_K^{i}]$, can be used as utterance identity representation. However, its dimension scales linearly with $K$. Instead, we decompose $\ve{w}_k^{i}$ into the product of a cluster-dependent loading matrix $\ve{T}_{k}$ and utterance identity vector $\ve{\omega}^{i}$ for more compact representation:
    \begin{equation}
    \mathbf{h}_{t}^{i}  \sim \sum_{k=1}^{K} z_{tk}^{i} \mathcal{N}\left(
    \boldsymbol{\mu}_{k} + \ve{T}_k\boldsymbol{\omega}^{i}, \ve{\Sigma}_k
    \right).
    \label{eq:ivector_fa}
    \end{equation}
    
    Specifically, we train a K-means model using the Transformer layer features to produce $\{\ve{\mu}_k\}$, \revision{which can be viewed as {\it content representations} of the speech.} Then, we run K-means to produce frame labels $\{z_{tk}^{i}\}$ and calculate $\{\ve{\Sigma}_{k}\}$ and cluster weight prior $\{\pi_{k}\}$ for the $K$ clusters, which we denoted as $\ve{\Phi}=\{\pi_k,\ve{\mu}_k,\ve{\Sigma}_k\ | k=1,\hdots,K\}$. With cluster parameters and frame labels $\{z_{tk}^{i}\}$, we only have one set of parameters $\{\ve{T}_k\}$ and one latent variable $\ve{\omega}^{i}$ left in the model, which is a problem that can be solved with the expectation-maximization (EM) algorithm.
    
    Given a sequence of frame-level features $\ve{H}^{i}=\{\ve{h}_{1}^{i},\hdots,\ve{h}_{T}^{i}\}$, the frames labels (alignments) $\ve{Z}^{i}=\{z^{i}_{tk}|t=1,\hdots,T;k=1,\hdots,K\}$, and cluster parameters $\ve{\Phi}$, we can use the EM algorithm to find $\ve{T}=\{\ve{T}_k|k=1,\hdots,K\}$. In the E-step, we compute the posterior of utterance identity $\ve{\omega}^{i}$:
    \begin{equation}
        p_{\ve{T}}\left(\ve{\omega}^{i}| \ve{H}^{i}; \ve{Z}^{i}, \ve{\ve{\Phi}}\right)   =
        \frac{
        \prod_{t=1}^{T} p_{\ve{T}}\left(\ve{h}^{i}_{t} | \ve{\omega}^{i};  \ve{z}^{i}_{ t\bullet } \right) 
        p\left(\ve{\omega}^{i}\right)
        }
        { \int \prod_{t=1}^{T} p_{\ve{T}}(\ve{h}^{i}_{t} | \ve{\omega}^{i};  \ve{z}^{i}_{ t\bullet } )  \text{d}\ve{\omega}^{i}
        },
        \label{eq:e-step}
    \end{equation}
    %%%%%%%%%%%%%%%%%%%%%%%%%%%%
    %%%%%%%%%%%%%%%%%%%%%%%%%%%%
    where $\ve{z}^{i}_{t \bullet }=\{{z}^{i}_{tk} \}_{k=1}^{K}$ and $p_{\ve{T}}\left(\ve{\omega}^{i}| \ve{H}^{i}; \ve{Z}^{i} ,\ve{\Phi}\right)$ is the probability distribution of $\ve{\omega}^{i}$ conditioned on $\ve{H}^{i}$ given $\ve{Z}^{i}$ and $\ve{\Phi}$. Because the alignments $\ve{Z}^{i}$ and the cluster parameters $\ve{\Phi}$ are fixed while optimizing the likelihood, we drop the dependency when expressing the posterior for simplicity.
    
    In the M-step, we choose the $\ve{T}$ that maximize the expected log-likelihood:
    \begin{equation}
    \operatorname*{arg\,max}_{\ve{T}} \sum_{i=1}^{I}\mathbb{E}_{p_{\ve{T}^{'}}(\ve{\omega}^{i}| \ve{H}^{i})}\left[\log p_{\ve{T}}\left(\ve{H}^{i}, \ve{\omega}^{i} \right)\right],
    \label{eq:m-step}
    \end{equation}
    where $\ve{T}^{'}$ is the loading matrix from the previous M-step (or randomly initialized).
    Eq.~\ref{eq:m-step} has a closed-form solution. 
    After the matrix $\ve{T}$ is found, the mean of the posterior $\mathop{\mathbb{E}}[\ve{\omega}|\ve{H}]$ is used as the utterance identity representation.
    \begin{equation}
    \label{eq:ivector}
        \mathop{\mathbb{E}}[\ve{\omega}|\ve{H}] = 
        (\ve{I} + \sum_{k}^{K}\ve{T}_k^{\text{T}}\ve{\Sigma}_{k}^{-1}\ve{T}_k)^{-1} \sum_{k}^{K}\ve{T}_k^{\text{T}}\ve{\Sigma}_{k}^{-1} \sum_{t} (\ve{h}_{t} - \ve{\mu}_{k}).
    \end{equation}
    
    {\bf Learning via gradient on ELBO}
    There are two limitations to learning matrix $\ve{T}$ using the EM algorithm. First, the EM algorithm limits the possibility of large-scale training. In Eq.~\ref{eq:m-step}, the loading matrix $\ve{T}$ is estimated using the whole training set, contrary to the stochastic update in modern DNN training. 
    Another disadvantage is the separation between the Transformer layers and the FA model during training, which prevents the possibility of joint optimization of the matrix $\ve{T}$ and Transformer layers' parameters $\ve{\theta}$.
    
    We aim to derive a learning rule that is amenable to stochastic updates and allows joint optimization of the FA model and the Transformer layers. As a latent variable model, the log-likelihood of our FA model can be written as \cite{bishop2006pattern,kingma2013auto}:
    \begin{equation}
    \log p_{\ve{T}}\left(\mathbf{H}^{i}\right)
    =D_\text{KL}\left(
        q(\ve{\omega}^{i}) \| p_{\ve{T}}(\ve{\omega}^{i}|\ve{H}^{i})
    \right)
    +\mathcal{L}_{\text{ELBO}}\left(
    \ve{H}^{i}  ; \ve{T}
    \right), 
    \label{eq:elbo}
    \end{equation}
    where $\mathcal{L}_{\text{ELBO}}\left(\ve{H}^{i} ; \ve{T} \right)$ is called the evidence lower bound (ELBO). $D_\text{KL}\left(
        q(\ve{\omega}^{i}) \| p_{\ve{T}}(\ve{\omega}^{i}|\ve{H}^{i})\right)$ is the KL-divergence between the approximate posterior $q(\ve{\omega}^{i})$ and true posterior $p_{\ve{T}}(\ve{\omega}^{i}|\ve{H}^{i})$. Minimizing KL or maximizing the ELBO can both increase the log-likelihood. In the case of our model, minimizing the KL is easy as the posterior of $\ve{\omega}$ is tractable, which gives rise to the E-step in Eq.~\ref{eq:e-step}. To optimize the ELBO, we need to re-write Eq.~\ref{eq:elbo} as:
    \begin{equation}
    \mathcal{L}_{\text{ELBO}}\left(\ve{H}^{i};\ve{T}\right)
    =\mathbb{E}_{q(\ve{\omega}^{i})}
    \left[
     -\log q(\ve{\omega}^{i})+ \log p_{\ve{T}}(\mathbf{H}^{i}, \ve{\omega}^{i})
    \right].
    \label{eq:elbo_original}
    \end{equation}
    Because we already know the closest ELBO to likelihood is when $q(\ve{\omega}^{i})$ equals to the posterior $p_{\ve{T}}\left(\boldsymbol{\omega}^{i} \mid \mathbf{H}^{i}\right)$, Eq.~\ref{eq:elbo_original} can be written as:
    \begin{equation}
    \mathbb{E}_{p_{\ve{T}^{'}}\left(\boldsymbol{\omega}^{i} \mid \mathbf{H}^{i}\right)}
    \left[
     -\log p_{\ve{T}^{'}}\left(\boldsymbol{\omega}^{i} \mid \mathbf{H}^{i}\right)+ \log p_{\ve{T}}(\mathbf{H}^{i}, \ve{\omega}^{i})
    \right],
    \label{eq:elbo_loss}
    \end{equation}
    where $\ve{T}^{'}$ is the loading matrix from the last update.
    We can see the first term is a constant with respect to $\ve{T}$.
    Therefore, the gradient of the lower-bound with respect to $\ve{T}$ is:
    \begin{equation}
        \frac{d\mathcal{L}_{\text{ELBO}}}{d\ve{T}} = \nabla_{\ve{T}} 
        \mathbb{E}_{p_{\ve{T}'}\left(\boldsymbol{\omega}^{i} \mid \mathbf{H}^{i}\right)}
         \left[ \log p_{\ve{T}}\left(\mathbf{H}^{i}, \ve{\omega}^{i}\right) \right].
         \label{eq:grad-to-T}
    \end{equation}
    The gradient with respect to the Transformer features $\frac{d\mathcal{L}_{\text{ELBO}}}{d\ve{H}^{i}}$ involves both terms in Eq.~\ref{eq:elbo_loss}:
    \begin{equation}
          \nabla_{\ve{H}^{i}}\mathbb{E}_{p_{\ve{T}^{'}}\left(\boldsymbol{\omega}^{i} \mid \mathbf{H}^{i}\right)}
    \left[
     -\log p_{\ve{T}^{'}}\left(\boldsymbol{\omega}^{i} \mid \mathbf{H}^{i}\right)+ \log p_{\ve{T}}(\mathbf{H}^{i}, \ve{\omega}^{i})
    \right].
    \end{equation}
    By applying the chain rule, we can obtain the gradient with respect to the Transformer parameters $\ve{\theta}$:
    \begin{equation}
        \frac{d\mathcal{L}_{\text{ELBO}}}{d\ve{\theta}} = \frac{d\mathcal{L}_{\text{ELBO}}}{d\ve{H}^{i}}
        \frac{d\ve{H}^{i}}{d\ve{\theta}}.
        \label{eq:grad-to-theta}
    \end{equation}
    Eq.~\ref{eq:grad-to-theta} shows that we can backpropagate the gradient of ELBO back to the Transformer layers.
    The total loss of our NFA model is:
    \begin{gather}
    \label{eq:final_loss}
            \sum_i \left(L_{m}(\ve{H}^{i}, \ve{Z}^{i}, {\cal M}) - \lambda \mathcal{L}_{\text{ELBO}} \left(\ve{H}^{i};\ve{T}\right)
        \right).
    \end{gather}
    Therefore, in addition to HuBERT's mask prediction and self-training, in each forward pass, we will compute the posteriors $p_{\ve{T}}\left(\boldsymbol{\omega}^{i} \mid \mathbf{H}^{i}\right)$ (Eq.~\ref{eq:e-step}) given a sequence of BERT features and frame labels produced by K-means. Then, we use the posteriors to evaluate the gradient with respect to $\ve{T}$ to update the loading matrix and the gradient with respect to BERT features $\ve{H}^i$ to update the SSL model parameters $\ve{\theta}$. Algorithm \ref{alg:join} summarizes the whole training procedure of our NFA.%We refer to the model trained this way as ELBO-NFA.
    \section{Experiments}
    In this section, we will evaluate the proposed NFA model's performance on three kinds of utterance-level speech tasks, namely speaker, emotion, and language recognition, by comparing it to SSL models such as wav2vec2.0, HuBERT, and WavLM. Note that the NFA can use both HuBERT and wav2vec2.0 architecture as long as frame labels are provided.
    \subsection{Tasks, Datasets, Baselines, and Implementation}
    \begin{table*}[t]
    \caption{Results on SUPERB and language identification tasks.} 
    \label{tab:superb}
    % \vskip 0.15in
    \begin{center}
    \begin{small}
    \begin{tabular}{l|c|c|c|c|c}
    \toprule
     Tasks & ASV & SD & SID & ER & LID  \\
            Metrics & EER $\downarrow$ & DER $\downarrow$ & Acc $\uparrow$ & Acc $\uparrow$ & Acc $\uparrow$ \\
            \midrule
            \textsc{wav2vec2.0 Large} \cite{superb}     & 5.65     & 5.62       & 86.14      & 65.64     & -     \\
                     \textsc{Supervised Finetuning \cite{wang2021fine}}     & 4.46    & -      & -     &  64.2    &     \\
                   \textsc{NFA (wav2vec2-based)}      & { 4.02}    & { 2.83 }     & { 96.3}     & { 73.4}    &     \\
                    %%%%%%%%%%%%%%%%
                    \midrule
            \textsc{HuBERT Large}  \cite{superb}    & 5.98     & 5.75  & 90.33  & 67.62     & -      \\
                   \textsc{WavLM Large}  \cite{wavlm}    & 3.77     & 3.24       & 95.49      & 70.62     & -      \\
                  \textsc{Supervised Finetuning HuBERT Large\cite{wang2021fine}}     & 2.36    & -      & -     &  72.7    &     \\
                                 \textsc{NFA (HuBERT-BASED)}  & {\bf 2.26} &  {\bf 1.84}  &  {\bf  98.1}  &  { 78.1}   & -      \\
    \midrule
      \textsc{Conformers} \cite{shor2022universal} & - &  - & - &  {\bf 79.2}   & -      \\
                        \midrule
           \textsc{wav2vec2-XLS-R}      & - & -  & - & -  & 80.4     \\
                 \textsc{ECAPA-TDNN}      & - & -  & - & -  & 84.9     \\
                    \textsc{NFA (XLS-R-BASED)}      &  - &  - &  - & - & {\bf  86.3}  \\
             \bottomrule
            \end{tabular}
    \end{small}
    \end{center}
    %\vskip -0.1in
    \end{table*}
    \begin{figure*}[!htp]
        \begin{center}
                \includegraphics[width=\textwidth]{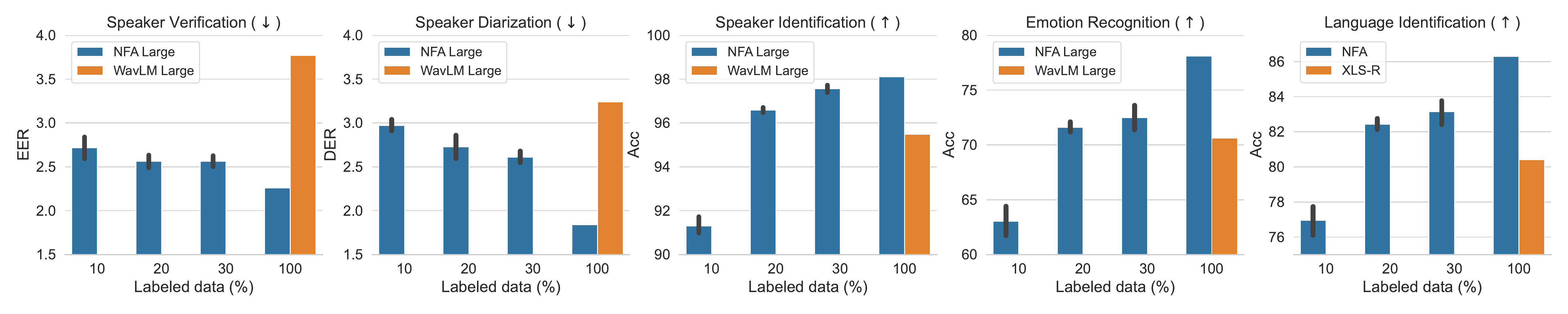}\\
            \end{center}
            \caption{Bar plots of SSL models' performance in low label-resource settings.}
        %	Note that there is no gradient (dashed arrow) flown back to the teacher model.}
            \label{fig:low-label-resource}
        \end{figure*}
    {\bf Speech Tasks and Datasets} The speech tasks that we will evaluate include:
    \begin{itemize}
    \item Automatic speaker verification (ASV or SV), speaker identification (SID), and speaker diarization (SD).
    We followed the SUPERB protocol \cite{superb} using the VoxCeleb1 \cite{vox1} training split to train the model and used the test split to evaluate speaker verification performance. Note that the reported ASV downstream model in \cite{superb} is a deep neural network \cite{snyder2018x} trained on SSL features \cite{superb}. The evaluation metric is equal error rate (EER) (the lower, the better).
    %Speaker identification is to identify a speaker from a pre-define set. 
    For speaker identification, we used the VoxCeleb1 train-test split provided by the SUPERB organizer. The evaluation metric is accuracy. For SID, the SUPERB downstream model is a linear classifier trained on averaged SSL features. 
    Speaker diarization is to segment and label a recording according to speakers. We followed the SUPERB protocol using the LibriSpeech \cite{librispeech} splits for training and evaluation. The SUPERB downstream model is a recurrent neural network. 
    The evaluation metric is diarization error rate (DER) (the lower, the better)
    \item Emotion recognition (ER). We used IEMOCAP \cite{IEMOCAP} dataset. Following the same protocol as SUPERB, we dropped the unbalance emotion classes to leave the neutral, happy, sad, and angry classes. The evaluation metric is accuracy. The SUPERB downstream model is a linear classifier trained on averaged SSL features. 
    \item Language identification (LID). Language identification is not included in the SUPERB benchmark. We included it because it is also an important utterance-level task. The dataset we used is the the Common Language dataset prepared by \cite{common-lang}, which includes 45 languages with 45.1 hours of recordings.  On average, each language has one-hour recordings.\footnote{\url{https://huggingface.co/datasets/common_language}} The downstream baseline is a linear classifier trained on averaged SSL features.
    \end{itemize}
    {\bf Pre-trained models} The pre-trained models we used in this paper include HuBERT \cite{hubert}, WavLM \cite{wavlm}, and wav2vec2-XLS-R \cite{XLS-R}. HuBERT and WavLM models were used in speaker and emotion evaluation. Because language identification requires models trained on multi-lingual data, wav2vec2-XLS-R was used.
    %To demonstrate the wide applicability of our methods, we include wav2vec 2.0 \cite{wav2vec2}, HuBERT \cite{hubert}, and WavLM \cite{wavlm} in our evaluation. Because language identification requires models trained on multi-lingual data, we use wav2vec2-XLS-R \cite{XLS-R}.
    
    {\bf Implementation details.} The HuBERT and Wav2vec2-based NFA models were trained on LibriSpeech using the model checkpoints provided by fairseq.
    % \footnote{\url{https://github.com/facebookresearch/fairseq/blob/main/examples/{hubert/README.md,wav2vec/README.md}}}
    The language identification NFA models were trained on the Common Language dataset using the XLS-R checkpoint.
    % \footnote{\url{https://github.com/facebookresearch/fairseq/blob/main/examples/wav2vec/xlsr/README.md}} 
    $\lambda$ in Eq.~\ref{eq:final_loss} is set to 0.01 for all models.
    After the optimization steps in Algorithm~\ref{alg:join} were done, we re-trained the loading matrix $\ve{T}$ for each task with EM using unlabeled task-related data. Other than specifically stated, the acoustic features were extracted from layer 6 for the base SSL models (HuBERT, WavLM, and Wav2Vec2-XLS-R) and layer 9 for the large SSL models.
    The number of clusters in K-means is 100, and the rank of loading matrix dimension is 300 for all NFA models. After utterance-level representations have been extracted using Eq.~\ref{eq:ivector}, we used the simple logistic classifier in sklearn \cite{pedregosa2011scikit} for SID, ER, and LID. For speaker verification, we used the PLDA backend. For SD, we used linear discriminant analysis (LDA) to reduce the dimension to 200 and then used agglomerative hierarchical clustering to produce speaker assignments. Note that all our downstream methods are linear models. 
    \subsection{SUPERB Experiments}
    In this section, we evaluate the NFA's performance on SUPERB tasks \cite{superb,wavlm}. Besides the standard speaker-related and emotion recognition, we also included language identification (LID) on Common Langue \cite{common-lang}. For LID, we followed the same protocol as other SUPERB tasks, i.e., the SSL models' weights were frozen, and only linear models were trained with labeled data without data augmentation. To give a better idea of the expected performance of each task in unrestricted settings, we also included the results using the fine-tuned SSL models on the ASV and ER tasks and the current best result in the Common Language dataset reported by other researchers. 
    
    \begin{figure}[!thp]
        \includegraphics[width=\columnwidth]{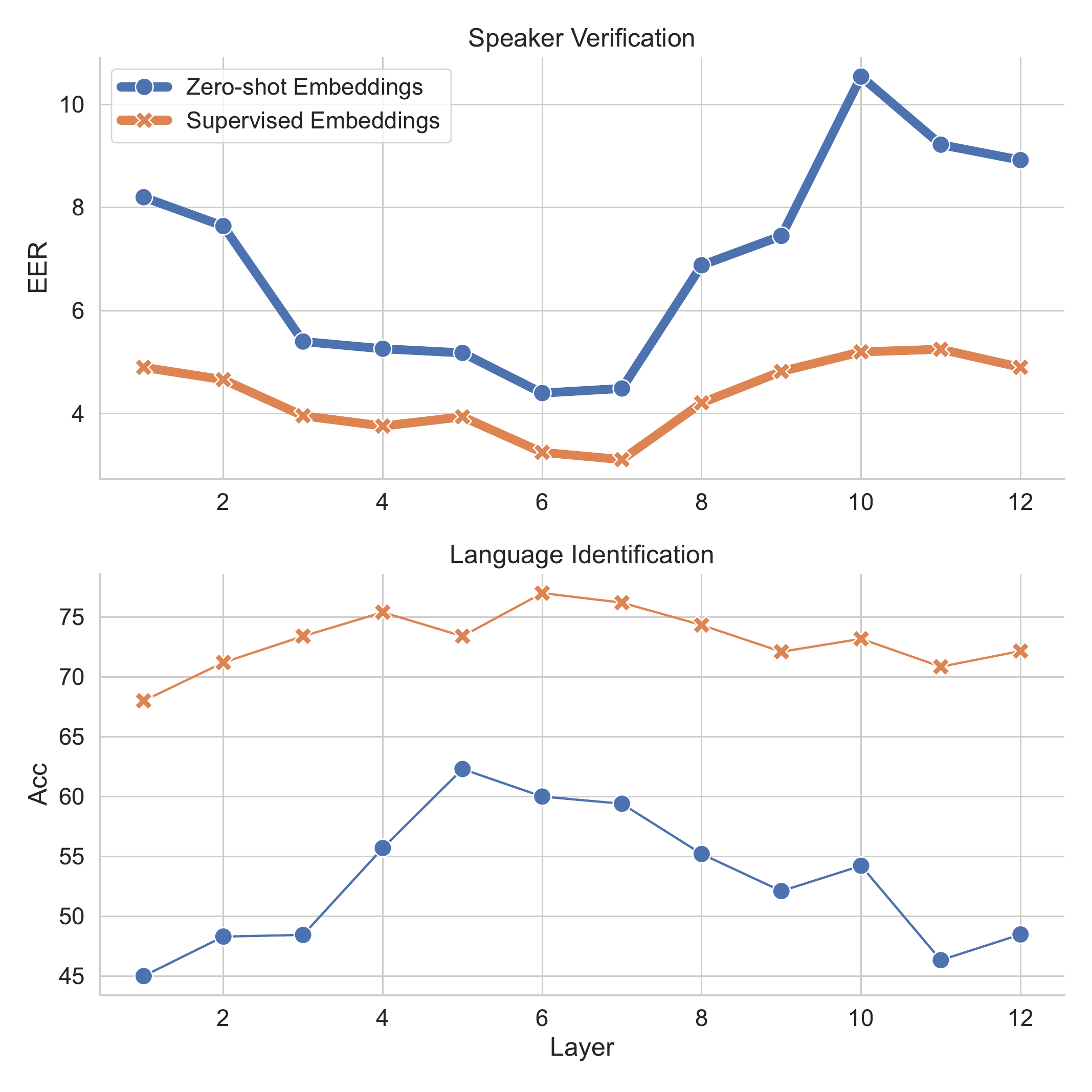}
        \caption{NFA embeddings' zero-shot performance on speaker verification and language ID.}
        \label{fig:layer-wise-zero-shot}
    \end{figure}
    The results are presented in Table~\ref{tab:superb}.
    \revision{
        As observed in the table, NFA significantly outperforms all SSL models across ASV, SD, SID, and LID. NFA performs only marginally worse than the self-supervised Conformer \cite{shor2020towards}, which has been specifically designed for utterance-level tasks.
    } In speaker verification, the relative EER reduction is 40\% when compared with the WavLM, the previous best model on utterance-level tasks. It is worth noting that WavLM's ASV baseline used a DNN network trained on the Transformer features, but we only use linear models. Our models even perform better than the fully fine-tuned models in \cite{wang2021fine} in both ASV and ER tasks. For LID, our XLS-R-based NFA performs better than the best-reported result on Common Language by SpeechBrain \cite{speechbrain}. 
    \subsection{Downstream Low Label-resource Experiments}
    One of the most attractive features of wav2Vec and HuBERT is their performance on low label-resource ASR. The resource efficiency of these models enables the potential development of many low label-resource languages and speech tasks where labeled data are hard to collect. In this section, we evaluate NFA performance in low label-resource settings. To this end, we divided the labeled dataset in the speaker recognition, emotion recognition, and language identification tasks into 10\%, 20\%, and 30\% subsets as low label-resource settings. For ASV, SD, SID, and ER, we extracted the embeddings from a large Hubert-based NFA model. For LID, we used the embeddings from the XLS-R-based NFA model. WavLM Large and XLS-R were used as performance references. 
    To reduce the performance deviation in the division, we ran each partition five times and reported the results. 
    The loading matrices in the NFA models were trained using the entire unlabeled dataset. 
    The results are presented in Figure~\ref{fig:low-label-resource}.
    
    We can see that even with only 10\%  of labeled data for the downstream models, NFA's performance in ER, SID, and LID is very close to the WavLM and XLS-R. For ASV and SD, our method already outperforms the WavLM models trained on fully labeled data. With 20\% labeled data, NFA already outperforms WavLM and XLS-R on all tasks. 
    This shows the high resource efficiency of our NFA models. 
    \subsection{Zero-Shot Speaker Verification}
    \label{sec:zero-shot}
    %%%%%%%%%%%%%%%%%%%%
    \begin{table}[t]
    \caption{Zero-shot speaker verification performance on different domains. The metric is the equal error rate.}
    \label{tab:sv-zero-shot}
    % \vskip 0.15in
    \begin{center}
    \begin{small}
    \begin{tabular}{l|c|c|c}
    \toprule
    Dataset & LibriSpeech & VoxCeleb & VOiCES  \\
            \midrule
                    \textsc{I-vector}     &  11.2      &   15.8    &            22.3 \\
                     \midrule
            \textsc{HuBERT }    &  28.7   &  32.1    &  34.5  \\
                    \textsc{NFA}   &  {\bf 3.98}   &  {\bf 9.32}  &         {\bf 12.32} \\
                     \midrule
                            \textsc{HuBERT Large}    &   30.21  & 26.88  &   37.45 \\
                    \textsc{NFA Large}   &  {\bf 2.87}   &  {\bf 7.92}  &         \bf{12.02} \\
             \bottomrule
            \end{tabular}
    \end{small}
    \end{center}
    %\vskip -0.1in
    \end{table}
    \begin{table*}
        \centering
    \begin{tabular}{l l c c c}
        \toprule
        Model Checkpoint & Optimization & ASV (EER) $\downarrow$& ER (ACC) $\uparrow$& Lang. ID (ACC) $\uparrow$\\
        \midrule
        HuBERT-Large & EM & 2.54\% & 73.4\% & - \\
        HuBERT-Large-NFA & Gradient & 2.26\% & 78.1\% & - \\
        Wav2vec-XLS-R & EM & - & - & 83.6\% \\
        Wav2vec-XLS-R-NFA & Gradient & - & - & 86.3\% \\
        \bottomrule
        \end{tabular}
        \caption{The performance of gradient-based learning versus EM.}
        \label{tab:sgd_vs_em}
    \end{table*}
    \begin{table}
        \centering
    \begin{tabular}{l c}
        \toprule
        Models & WER \\
        \midrule
        Base HuBERT & 6.42 \\
        Base NFA & 6.31 \\
        Large HuBERT & 3.62 \\
        Large NFA & 3.66 \\
        \bottomrule
        \end{tabular}
        \caption{ASR performance on LibriSpeech clean subset.}
        \label{tab:asr}
    \end{table}
    %\subsection{Implication}
    % 1/19 stop here
    In Figure~\ref{fig:umap}, we observe that by clustering and aligning the Transformer features, speaker information can be revealed. This is all done without labeled data. But how discriminative these unsupervised learned embeddings are?
    We will evaluate NFA embeddings' zero-shot performance quantitatively in this section. 
    Specifically, we evaluated NFA models on zero-shot speaker verification. After we extracted the utterance-level representations using Eq.~\ref{eq:ivector}, we directly used cosine similarity to obtain verification scores without any supervised training (the models were never given speaker information). 
    We evaluated the performance on (1) LibriSpeech, which is considered in-domain data as HuBERT and NFA were trained on this dataset \cite{librispeech,hubert}, (2) Voxceleb1-test, a popular speaker verification dataset \cite{vox1}, and (3) VOiCES \cite{nandwana2019voices}, a dataset used to evaluated speaker verification robustness against noise and room reverberation. 
    As a comparison, we also included i-vector \cite{ivector} and averaged Transformer features (HuBERT rows in Table~\ref{tab:sv-zero-shot}) as baselines. 
    
    The results are presented in Table~\ref{tab:sv-zero-shot}. Without supervision, simple averaging the Transformer features cannot produce useful speaker representations. It even performs worse than i-vector, a non-DNN approach. NFA embeddings, however, achieve an EER of 3.98\% on LibriSpeech without any supervised training. This suggests that during self-supervised learning, the model has already learned to differentiate speakers, \revision{which also empirically demonstrates that the NFA model can disentangle speaker information from the content information.}
    However, when evaluated on VoxCeleb1 and VOiCES, the performance of zero-shot SV dropped significantly. This may be because VoxCeleb1 and VOiCES are real-world speech datasets containing spontaneous speech and environmental noise. NFA and HuBERT were pre-trained on a read speech dataset. The domain discrepancy in SSL models can have a significant impact on the downstream tasks, as mentioned in \cite{SSL-domain}. Another interesting observation is that scaling the model size improves the zero-shot SV performance, as shown when using HuBERT Large and NFA large models.
    \subsection{Layer-wise Representation Evaluation}
    Because our NFA models show excellent zero-shot performance, we can use them to evaluate the discrimination power from each Transformer layer before supervised learning is applied. We extracted the acoustic features from Layer 1 to Layer 12 of the Transformer in the NFA model to conduct zero-shot speaker verification and language identification. For language identification, we used top-1 accuracy as the metric. Then, we used the labeled data to train an LDA on top of NFA embeddings to compare the results. The results are presented in Figure~\ref{fig:layer-wise-zero-shot}.
    
    The blue lines in Figure~\ref{fig:layer-wise-zero-shot} show that under zero-shot settings, both speaker and language discriminative abilities increase from Layer 1 up to Layer 6. Then, the features from the deeper layers have poorer performance. This is largely consistent with the supervised baselines (orange lines), with Layer 7 obtaining the lowest speaker verification error and Layer 6 having the highest language identification top-1 accuracy in supervised settings. This shows that our NFA models' zero-shot performance can be a reliable predictor of supervised performance.
    \subsection{Gradient-based Learning Versus EM}
    To assess whether gradient-based learning has an edge over the Expectation-Maximization (EM) method, we extracted HuBERT features and separately trained a factor analysis model using EM. The results are displayed in Table~\ref{tab:sgd_vs_em}. We observe that gradient-based optimization consistently outperforms EM-based I-vector trained on HuBERT features. This suggests that jointly training the NFA model with the SSL model can yield more potent feature representations than training the two modules independently.
    \subsection{Impact on ASR}
    The ultimate goal of a self-supervised learning (SSL) speech model is to utilize a single backbone model for all downstream tasks. Consequently, it's critical that the NFA model does not compromise performance on content-based tasks such as ASR. To ensure this, we compared the performance of the NFA and the large NFA model against HuBERT on the LibriSpeech clean subset. The results, as shown in Table~\ref{tab:asr}, demonstrate that the NFA and large NFA models perform on par with HuBERT. This confirms that our NFA model does not sacrifice performance on content-based tasks.
 \section{Conclusions}
    In this paper, we proposed a novel self-supervised speech model for utterance-level speech tasks. Instead of using frame-wise discrimination loss alone, we introduced an utterance-level learning objective based on factor analysis and feature disentanglement. Through extensive experiments, we demonstrate that our NFA model can significantly improve SSL models' performance on utterance-level discriminative tasks without supervised fine-tuning. 
    The zero-shot and low label-resource experiments also show the data efficiency of our approach, which to the best of our knowledge, has yet been shown for utterance-level tasks. This can significantly benefit the utterance-level speech classification tasks where labeled data is hard to obtain, such as speaker recognition for low label-resource languages \cite{thanh2021deep}, depression speech detection \cite{ma2016depaudionet}, children speech processing \cite{shahnawazuddin2021children}, speech disorder diagnosis \cite{alhanai2017spoken}, and classifying intelligibility for disordered speech \cite{venugopalan2021comparing}.
    
    Our findings also shed some insights into speech SSL learning itself. Currently, the frame-wise discriminative SSL models are often thought of as acoustic unit discovery models. Little has been considered for utterance-level identity discovery such as speaker information in self-supervised learning. As we show in Section~\ref{sec:zero-shot}, SSL can perform very well on speaker verification with supervision, which suggests speaker-related information is also discovered during the self-supervised learning stage. This is encouraging as it shows that SSL learning can discover multiple hidden information in the speech that can benefit a wide range of speech tasks.
    
    \revision{
    A significant limitation of the NFA model lies in its performance with out-of-domain data. As observed in Section~\ref{sec:zero-shot}, NFA's performance significantly deteriorates when evaluated on out-of-domain data. This observation underscores the persistent challenge of achieving robust zero-shot performance in SSL models.
    Another limitation of NFA pertains to the types of signals it can effectively disentangle. While the NFA model showcases impressive feature disentanglement capabilities across several utterance-level tasks, it's worth noting that it does not disentangle different types of utterance-level information from one another. For instance, it does not separate speaker information from emotional states. For such nuanced tasks, we continue to rely on downstream models to achieve this level of disentanglement. In future research, we intend to explore methodologies that could disentangle different types of utterance-level information during the self-supervised learning stage.
    }
    %\section*{Acknowledgements}
    %
    %\textbf{Do not} include acknowledgements in the initial version of
    %the paper submitted for blind review.
    %
    %If a paper is accepted, the final camera-ready version can (and
    %probably should) include acknowledgements. In this case, please
    %place such acknowledgements in an unnumbered section at the
    %end of the paper. Typically, this will include thanks to reviewers
    %who gave useful comments, to colleagues who contributed to the ideas,
    %and to funding agencies and corporate sponsors that provided financial
    %support.

    \bibliography{./example_paper}
    \bibliographystyle{icml2023}

%%%%%%%%%%%%%%%%%%%%%%%%%%%%%%%%%%%%%%%%%%%%%%%%%%%%%%%%%%%%%%%%%%%%%%%%%%%%%%%
%%%%%%%%%%%%%%%%%%%%%%%%%%%%%%%%%%%%%%%%%%%%%%%%%%%%%%%%%%%%%%%%%%%%%%%%%%%%%%%
% APPENDIX
%%%%%%%%%%%%%%%%%%%%%%%%%%%%%%%%%%%%%%%%%%%%%%%%%%%%%%%%%%%%%%%%%%%%%%%%%%%%%%%
%%%%%%%%%%%%%%%%%%%%%%%%%%%%%%%%%%%%%%%%%%%%%%%%%%%%%%%%%%%%%%%%%%%%%%%%%%%%%%%
\newpage
\appendix
\onecolumn
\section{You \emph{can} have an appendix here.}

You can have as much text here as you want. The main body must be at most $8$ pages long.
For the final version, one more page can be added.
If you want, you can use an appendix like this one, even using the one-column format.
%%%%%%%%%%%%%%%%%%%%%%%%%%%%%%%%%%%%%%%%%%%%%%%%%%%%%%%%%%%%%%%%%%%%%%%%%%%%%%%
%%%%%%%%%%%%%%%%%%%%%%%%%%%%%%%%%%%%%%%%%%%%%%%%%%%%%%%%%%%%%%%%%%%%%%%%%%%%%%%

\end{document}